\documentclass[12pt]{iopart}

\usepackage{graphicx}

\begin{document}

\title[Nano-optical Yagi-Uda antenna]{Design parameters
for a nano-optical\\ Yagi-Uda antenna}

\author{Holger F. Hofmann, Terukazu Kosako and Yutaka Kadoya}

\address{Graduate School of Advanced Sciences of Matter, Hiroshima University,
Kagamiyama 1-3-1, Higashi Hiroshima 739-8530, Japan}

\begin{abstract}
We investigate the possibility of directing optical emissions
using a Yagi-Uda antenna composed of a finite linear array
of nanoparticles. The relevant parameters characterizing the
plasma resonances of the nanoparticles are identified and
the interaction between the array elements is formulated
accordingly. It is shown that the directionality of the optical
emission can be observed even in the presence of
non-negligible absorption losses in the nanoparticles.
We conclude that a finite array of gold nanorods may be
sufficient for the realization of a working nano-optical
Yagi-Uda antenna.
\end{abstract}


\ead{h.hofmann@osa.org}

\maketitle

\section{Introduction}

Nanoscience is essentially the art of applying engineering
principles at scales where some of our conventional assumptions
cease to be valid. One interesting example is the
control of light using the plasma resonances of metallic
nanoparticles. The optical response of metal
nanoparticles is well described by classical electrodynamics,
with the material properties represented by the frequency
dependent dielectric constant. It is therefore possible to
apply ideas from electrical engineering directly to the
design of nano-optical devices \cite{Eng05}. However, the optical
response of metals is quite different from the metallic conductivity
observed low frequencies. One well researched consequence of this
difference is the observation of resonant excitations known as
surface plasmons in nanoparticles much smaller than the optical
wavelength \cite{Mai05}. Such resonances can be used to
enhance and direct the spontaneous emission of light by
single molecules, quantum
dots, or similar point-like light sources, just like an
appropriately designed resonant antenna can enhance and direct
radio emissions \cite{Ger00,Gre05}. Recently,
this idea has been combined with
the principles of antenna design known from radio and microwave
technology, resulting in the experimental realizations of
nano-optical equivalents of half-wave and bow-tie antennas
\cite{Sch05,Muh05,Far05,Cub06}. Based on these advances in the fabrication of
nano-optical antennas, it seems only natural to ask whether
the same principles can be used to also achieve the highly
directional emission of radio antennas in the optical regime.

At radio frequencies, high directionality can be obtained
by conventional Yagi-Uda antennas consisting of a reflector and
several directors arranged in line around the feed. In the
optical regime, the same kind of antenna structure could be
realized by a linear array of nanoparticles.
In our recent work, we therefore pointed out the possibility
of using a nano-optical Yagi-Uda antenna made of metal nanorods
to direct the optical emission from systems such as molecules,
quantum dots or optical semiconductors \cite{Kos06}. A similar
proposal based on concentric core-shell nanoparticles
was developed independently by Jingjing Li et al. at the
University of Pennsylvania \cite{Li07}.

Since there appears to be a growing interest in this
kind of device structure, it may be useful to formulate the
essential theory of Yagi-Uda antennas in an accessible manner.
In the following, we therefore develop a simple theory of the
nano-optical Yagi-Uda antenna that allows us to identify the
relevant properties of the nanoparticles and their effect on the
directivity of the emission. To do so, we have to consider both the
basic electrodynamics of near field coupling between dipoles at
distances close to the wavelength of their emission
(section \ref{sec:coupling}), and the interaction between the fields
and the material properties inside each nanoparticle
(section \ref{sec:particle}). The result allows us to estimate the
relevant antenna parameters for gold nanorods embedded in a
glass substrate (section \ref{sec:gold}). We then apply the
near field coupling equations of section \ref{sec:coupling}
to derive the emission pattern of Yagi-Uda antenna arrays
(section \ref{sec:array}) and use the results to optimize
the detuning and the array spacings of a five element
antenna array (section \ref{sec:optimize}).
Finally, we investigate the effect of non-negligible absorption
losses on the emission pattern (section \ref{sec:emission})
and show that the directivity of the emission is sufficiently
robust against the losses expected for the gold nanorods
described in section \ref{sec:gold}.

\section{Near field interactions and radiative damping}
\label{sec:coupling}

In order to have a maximal effect on the emission pattern,
antenna elements must be placed at distances comparable to the
wavelength $\lambda$ of the emitted radiation. Therefore,
the interaction between the elements is described by the
electric field in the transitional range between the quasi-static
dipole near field and the far field.
For an oscillating dipole $d$ emitting radiation of wavelength
$\lambda$ into a homogeneous medium of dielectric constant
$\epsilon_{\mbox{\small med.}}$, the field pattern is determined by
Hertz's solution of electric dipole radiation.
At a distance of $r$ perpendicular to the orientation of the
oscillating dipole, the electric field can then be given as
\[
E_z(x;A) = -\frac{3}{2} A \frac{1}{x^3}(i + x - i x^2) \exp(i x),
\]
where
\begin{equation}
\label{eq:inter}
A = - i \frac{4 \pi^2 d}{3 \epsilon_{\mbox{\small med.}} \lambda^3}
\hspace{1cm} \mbox{and} \hspace{1cm}
x = 2 \pi \frac{r}{\lambda}.
\end{equation}
The parameter $A$ expresses the dipole in terms of its
radiation damping field (that is, its self-interaction). This
representation of the dipole greatly simplifies the representation
of polarizability. For a resonant oscillator without
internal losses, the incoming field $E_{\mbox{\small in}}$
is exactly compensated by the radiation damping field of the
induced dipole. Therefore, the dipole of a resonant loss free
nanoparticle is determined by $A=-E_{\mbox{\small in}}$.
In the more general case of a finite detuning and
additional absorption losses, the scaled response function can be
expressed as
\begin{equation}
\label{eq:resp}
A = - \frac{E_{\mbox{\small in}}}{1 + \gamma +i \delta},
\end{equation}
where $\delta$ and $\gamma$ describe the effects of detuning
and of absorption losses as ratios of the associated fields
to the radiation damping field of the dipole.

Equations (\ref{eq:inter}) and (\ref{eq:resp}) describe the
coupling effects in any linear dipole array at any wavelength.
The material properties of the system are summarized by
the two parameters $\gamma$ and $\delta$, describing the losses
and the spectral properties of the array elements.
The possibility of designing a nano-optical Yagi-Uda antenna
therefore depends mainly on the ability to control these two
parameters of the individual nanoparticles.

\section{Optical properties of the nanoparticles}
\label{sec:particle}

A major difference between nano-optical antennas and radio frequency
antennas is that the plasma resonance of a metallic nanoparticle is
nearly independent of size. In principle, the elements of an
antenna array can therefore be much smaller than the half
wavelength size of classical antenna designs \cite{Muh05}.
The reason for this size independence is
that the role of the inductance of the classical antenna is
taken over by the mass of the free electrons in the plasma oscillation.
This effect of the electron mass is reflected by the negative
real part of the dielectric constants of metals at optical
frequencies, which converts the conventional capacitive effect of
polarizability into an inductive effect \cite{Eng05}.
The polarization $P$ of a nanoparticle by an incoming field
$E_{\mbox{\small in}}$ can thus be described in close analogy to the
response of an electrical circuit to the application of a voltage.
Specifically, the response can be separated into the response of the
material inside the particle described by the complex dielectric
constant $\epsilon_{\mbox{\small particle}}=
\epsilon_r \epsilon_{\mbox{med.}}$, the shape dependent field of
the surface charges accumulated at the interface between the
nanoparticle and the surrounding medium, and the radiation
damping field caused by the emission of the total dipole
$d=P V$, where $V$ is the volume of the nanoparticle.
The relation between incoming field and polarization then reads
\begin{equation}
\label{eq:circuit}
E_{\mbox{in}} = \underbrace{\frac{1}{(\epsilon_r-1)
\epsilon_{\mbox{\small med.}}} P }_{\mbox{\parbox{1.5cm}{\small material \\ (inductive)}}} +
\underbrace{\frac{N}{\epsilon_{\mbox{\small med.}}} P }_{\mbox{\parbox{1.5cm}{\small
shape \\ (capacitive)}}} - i
\underbrace{\frac{4 \pi^2 V}{3 \epsilon_{\mbox{\small med.}}
\lambda^3} P }_{\mbox{\parbox{1.5cm}{\small radiation \\ (resistive)}}}.
\end{equation}
Here, the capacitive effect of the surface
charge is given in terms
of the depolarization factor $N$, which is defined in close
analogy to the demagnetization factor in the magnetic response
of a particle as the factor by which a polarization $P$ weakens
the electric field $E$ inside a body of a given shape.
In its more conventional form, the polarization of
a nanoparticle is given as
\begin{equation}
\label{eq:spec}
P = \frac{(\epsilon_r-1)\epsilon_{\mbox{\small med.}}
E_{\mbox{in}}}{1+N(\epsilon_r-1)-i (\epsilon_r-1) R}, \hspace{0.5cm}\mbox{where}
\hspace{0.5cm} R=\frac{4 \pi^2 V}{3 \lambda^3}.
\end{equation}
This equation is equivalent to the antenna response equation (\ref{eq:resp}) in section \ref{sec:coupling}, where $P$ is
expressed in terms of the radiation damping field,
$A=-i R P /\epsilon_{\mbox{\small med.}}$.
We can therefore identify the parameters
$\gamma$ and $\delta$ in eq.(\ref{eq:resp}) with the corresponding
terms in eq.(\ref{eq:spec}) to find the relative losses $\gamma$
and the effective detuning $\delta$ of a nanoparticle based on its dielectric constant $\epsilon_r=\epsilon_{\mbox{\small particle}}/
\epsilon_{\mbox{med.}}$, its volume $V$ (given in terms of the
scaled parameter $R$), and its depolarization factor $N$.
The results read
\begin{eqnarray}
\label{eq:gamma}
\gamma &=& \frac{1}{R} \frac{\mbox{Im}(\epsilon_r)}{|\epsilon_r-1|^2}
\\
\label{eq:delta}
\delta &=& \frac{1}{R} \left(N+\frac{\mbox{Re}(\epsilon_r-1)}{|\epsilon_r-1|^2} \right).
\end{eqnarray}
Using these relations, we can now determine whether a
specific type of nanoparticle is suitable for the construction of
an antenna array.

\section{Shape and size dependences for gold nanoparticles embedded in glass}
\label{sec:gold}

\begin{table}
\caption{\label{gold} Optical properties of gold nanorods embedded in
a glass substrate with a dielectric constant of $\epsilon_{\mbox{\small med.}}=2.3$.}
\begin{indented}
\lineup
\item[]
\begin{tabular}{lllll}
\br
Wavelength & $\epsilon_{\mbox{\small particle}}$ && Wavelength &
\\
in vacuum & of gold \cite{Joh72} && in glass, $\lambda$ & $\epsilon_r-1$
\\ \mr
582 nm & -8.11+i 1.66 && 384 nm & -4.53 + i 0.72
\\
617 nm & -10.7+i 1.37 && 407 nm & -5.65 + i 0.60
\\
660 nm & -13.6+i 1.04 && 435 nm & -6.91 + i 0.45
\\
705 nm & -16.8+i 1.07 && 465 nm & -8.30 + i 0.47
\\
756 nm & -20.6+i 1.27 && 499 nm & -9.96 + i 0.55
\\
821 nm & -25.8+i 1.63 && 541 nm & -12.2 + i 0.71
\\
\br
\end{tabular}
\end{indented}
\end{table}
One simple and convenient way to realize a nano-optical Yagi-Uda
antenna may be to use gold nanoparticles embedded in a glass
substrate. The optical properties of gold nanoparticles are well
known and the dielectric constants at different frequencies is
readily available \cite{Joh72}. We have used this data to determine
the optical properties of gold nanoparticles in a glass substrate
with a dielectric constant of $\epsilon_{\mbox{\small med.}}=2.3$.
Table \ref{gold} shows the wavelengths inside the glass and the
values of $\epsilon_r-1$ that determine the dependence of the loss
and detuning parameters on the size and shape of the nanoparticles.

The main limitations on realizing a nano-optical antenna array with
gold particles arises from the absorption losses represented by the
imaginary part of the dielectric constant $\epsilon_r$ describing
the optical response of gold relative to the surrounding glass.
According to eq.(\ref{eq:gamma}), the relevant loss parameter
$\gamma$ is given by the ratio of material properties and the
volume of the nanoparticle. It is therefore possible to compensate
the absorption losses of the material by making the volume of
the nanoparticle larger. The volume at which the loss parameter
$\gamma$ becomes one may be used as a reference point for an
estimate of the necessary particle size. Table \ref{losses}
illustrates this volume dependence of the losses by giving the
product of losses and scaled volume $\gamma R$, as well as the
volume at which the loss parameters is $\gamma = 1$ in terms of
the cubic wavelength in glass.
\begin{table}
\caption{\label{losses} Volume dependence of losses for
gold nanoparticles embedded in a glass substrate.}
\begin{indented}
\lineup
\item[]
\begin{tabular}{lll}
\br
Wavelength in &  &
\\
vacuum (glass) & $\gamma R$ & $V(\gamma=1)$
\\ \mr
582 nm ($\lambda=$  384 nm) &
$3.42 \times 10^{-2}$ & $2.6 \times 10^{-3} \lambda^3$
\\
617 nm  ($\lambda=$  407 nm) &
1.86 $\times 10^{-2}$ & $1.4 \times 10^{-3} \lambda^3$
\\
660 nm  ($\lambda=$  435 nm) &
1.20 $\times 10^{-2}$ & $9.1 \times 10^{-4} \lambda^3$
\\
705 nm  ($\lambda=$  465 nm) &
6.80 $\times 10^{-3}$ & $5.2 \times 10^{-4} \lambda^3$
\\
756 nm  ($\lambda=$  499 nm) &
5.53 $\times 10^{-3}$ & $4.2 \times 10^{-4} \lambda^3$
\\
821 nm  ($\lambda=$  541 nm) &
4.75 $\times 10^{-3}$ & $3.6 \times 10^{-4} \lambda^3$
\\
\br
\end{tabular}
\end{indented}
\end{table}
The results indicate that the dimensions of the antenna elements
should be of the order of a tenth of the wavelength inside the
medium. This is still small enough to justify the approximation
as a local dipole oscillator, while also being large enough to
permit a definition of the device structure by standard lithographic
techniques.

Once the size has been fixed, the precise detuning can be determined
by varying the shape. The relation between the detuning parameter
$\delta$ and the depolarization factor $N$ is given by eq.
(\ref{eq:delta}). The depolarization factor can be adjusted by
varying the aspect ratio of the nanoparticle, defined as the
ratio of the length $a$ along the direction of the electric field
and the perpendicular width $b$. For the case of an ellipsoid,
analytical formulas are known from the study of demagnetization
\cite{Sun05}, and these can be used to estimate the approximate
aspect ratios for nanorods tuned to a specific frequency. However,
the exact relation between the depolarization factor $N$ and the
aspect ratio may depend on details of the shape, so it may be best
to obtain it experimentally for any given geometry.
\begin{table}
\caption{\label{tuning} Depolarization factors for resonant antenna elements and the corresponding aspect ratios for the case of
ellipsoids.}
\begin{indented}
\lineup
\item[]
\begin{tabular}{lll}
\br
Wavelength in & & Aspect
\\
vacuum (glass) &  $N (\delta=0)$ & ratio $a/b$
\\ \mr
 582 nm ($\lambda=$  384 nm) & 0.215 (about 1/5) & 1.63
\\
617 nm  ($\lambda=$ 407 nm) & 0.175 (about 1/6) & 1.99
\\
660 nm  ($\lambda=$ 435 nm) & 0.144 (about 1/7) & 2.37
\\
705 nm ($\lambda=$ 465 nm) & 0.120 (about 1/8) & 2.76
\\
756 nm ($\lambda=$ 499 nm) & 0.100 (exactly 1/10) & 3.21
\\
821 nm ($\lambda=$ 541 nm) & 0.082 (about 1/12) & 3.75
\\
\br
\end{tabular}
\end{indented}
\end{table}
Table \ref{tuning} shows the depolarization factor necessary
to establish resonance at each of the frequencies given in
table \ref{gold}, together with the corresponding aspect ratio
for an ellipsoid.

\begin{figure}
\begin{center}
\scalebox{1.0}[1.0]{\includegraphics{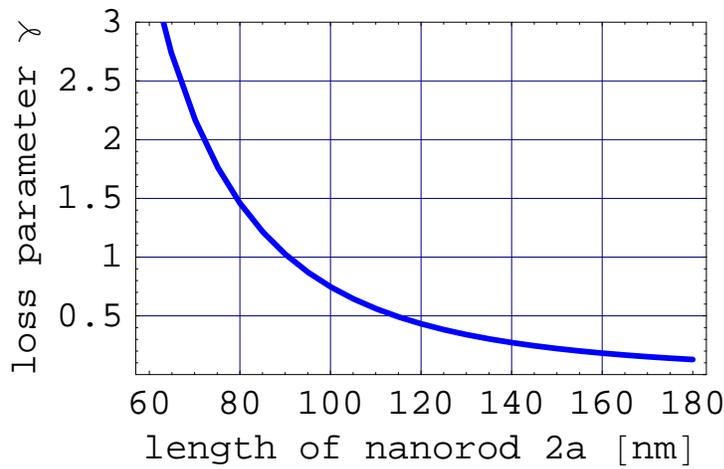}}
\end{center}
\caption{\label{length}
Length dependence of the loss parameter $\gamma$ for an elliptical
gold nanorod of aspect ratio 2.76 at its resonant wavelength of
705 nm.
}
\end{figure}

We can now get a practical idea of the relation between the
loss parameter $\gamma$ and the size of the nanoparticles used
to construct the antenna by relating $\gamma$ to the length of
an ellipsoid with the appropriate aspect ratio. Fig.\ref{length}
shows the dependence of losses on length for a particle
of aspect ratio 2.76 at its resonant wavelength of 705 nm.
A loss parameter of $\gamma=0.5$ is obtained for particles
with a length of $2a=115 \mbox{nm}$, corresponding to
$0.25 \lambda$ inside glass ($\lambda=465 \mbox{nm}$).
To obtain a loss parameter of $\gamma=0.1$, the length of the
gold nanorods must be increased to $2a=0.42 \lambda$, or about
195 nm. For gold nanoparticles sufficiently shorter than the
half wavelength of classical antennas, it is therefore unrealistic
to neglect the effects of losses on the performance of the
antenna. However, losses around $\gamma=0.5$ are still reasonably
low, so particles of about 100 nm may be suitable for use in
an antenna array.

In order to construct a Yagi-Uda antenna, it is also necessary
to control the detuning parameter $\delta$ of the elements used
as directors. As eq.(\ref{eq:delta}) shows, the detuning
caused by a change of $\Delta N$ in the detuning factor is
equal to $\Delta N/R$. We can therefore determine the change in
depolarization factor required to achieve the intended detuning
$\delta$ by multiplying $\delta$ with $R$. For example, an antenna
element operating at 705 nm with losses of $\gamma=0.5$
can be detuned to $\delta=1$ by increasing the depolarization
factor to $0.120+R=0.136$. This depolarization factor corresponds
to an aspect ratio of about 2.5 for ellipsoids. Using the
volume corresponding to losses of $\gamma=0.5$ at 705 nm,
this aspect ratio defines a length of $2a=107$ nm for the ellipsoid.
Roughly speaking, a reduction of 7\% in length thus corresponds
to an increase of one in the detuning parameter $\delta$.

\section{Coupled dipole theory of the antenna array}
\label{sec:array}

The operating principle of the Yagi-Uda antenna is based on
the observation that capacitively detuned antenna elements
effectively guide the emissions from a neighboring feed
towards their direction, while resonant or inductively detuned
elements always reflect the emissions back towards the feed.
A Yagi-Uda antenna therefore consists of an array of several
directors placed in front of the feed, and one or two reflectors
behind the feed. At radio frequencies, the description of a
Yagi-Uda antenna is usually complicated by the fact that the
length of the antenna rods exceeds the distance between
antenna elements. It is then unrealistic to assume a simple
dipole coupling interaction, such as the one given by
eq. (\ref{eq:inter}). However, in the case of nanoparticles,
the dipole coupling approximation is usually sufficient to
describe interactions in an array \cite{Mar93,Zou06}. A nano-optical
Yagi-Uda antenna can therefore be described in terms of the
fundamental physics of an interacting dipole chain using the
formalism developed in section \ref{sec:coupling}.

In the following, we will consider a number of
antenna elements at positions $x_n$ along the antenna axis
around a feed at $x=0$.
As in eq. (\ref{eq:inter}), all distances
are given in units of $\lambda/(2 \pi)$. The feed is a light
emitting system such as a molecule or quantum dot, oscillating at
a well defined frequency. The dipole amplitude of the feed oscillation
is given by $A_F$. Each passive antenna elements responds to this
feed oscillation with a dipole amplitude $A_n$ proportional to
$A_F$. The precise interactions are defined by eqs.(\ref{eq:inter})
and (\ref{eq:resp}). They form a system of $N$ linear equations,
where $N$ is the number of passive elements of the antenna array,
\begin{equation}
\label{eq:array}
A_n = \frac{1}{1+\gamma_n+i\delta_n}\left(E_z(|x_n|;A_F) +
\sum_{m\neq m} E_z(|x_n-x_m|;A_m)\right).
\end{equation}
By solving this system of linear equations for a particular
configuration of antenna elements, we can determine the
ratio between the dipole amplitudes $A_n$ and the feed
amplitude $A_F$. From these amplitudes, we can then derive the
far field emission pattern. In the plane orthogonal to the
dipoles, this pattern is given by
\begin{equation}
I(\theta)/I_F=\left(|1+\sum_n \frac{A_n}{A_F}
\exp(-i x_n \cos(\theta))|^2\right),
\end{equation}
where $\theta$ is the angle between the axis of the antenna array
and the emission direction and $I_F$ is the intensity of emission
from a single dipole oscillating at the feed amplitude $A_F$.

\section{Basic structure of a five element Yagi-Uda antenna}
\label{sec:optimize}

\begin{figure}
\begin{center}
\scalebox{1.60}[1.60]{\includegraphics{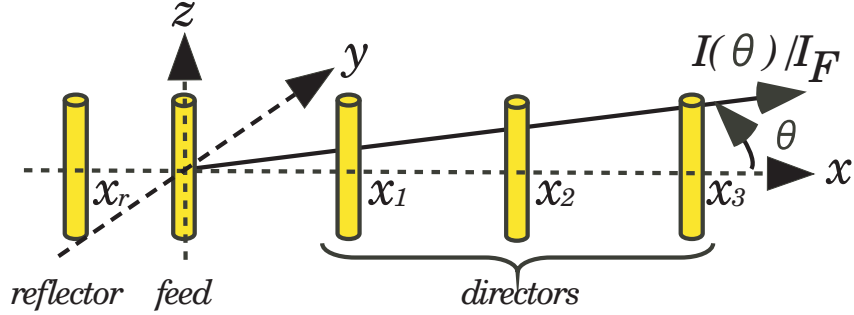}}
\end{center}
\caption{\label{YU}
Sketch of a five element Yagi-Uda antenna, indicating the positions
of the reflector, feed, and directors, and the angle of emission
$\theta$ in the plane orthogonal to the dipoles of the antenna
elements.}
\end{figure}

Fig. \ref{YU} illustrates the basic structure of a five element
Yagi-Uda antenna. It consists of a resonant reflector at $x_r<0$
($\delta_r=0$), the feed at $x=0$, and three equally spaced
directors at $x_1=x_a$, $x_2=2 x_a$, and $x_3=3 x_a$, capacitively
detuned by $\delta=\delta_1=\delta_2=\delta_3$.
A typical choice of parameters for radio frequency antennas is
$x_r=-1.4$ corresponding to a reflector distance of $0.22 \lambda$,
$x_a=2$ corresponding to a director spacing of $\lambda/\pi$,
and a detuning parameter of $\delta=1$. However, the exact choice
of parameters is not very critical, making a Yagi-Uda antenna
very robust against misalignments of its elements.

\begin{figure}
\begin{center}
\begin{picture}(360,400)
\put(10,380){\makebox(20,20){(a)}}
\put(10,180){\makebox(20,20){(b)}}
\put(40,200){\makebox(280,200){
\scalebox{1.05}[1.05]{\includegraphics{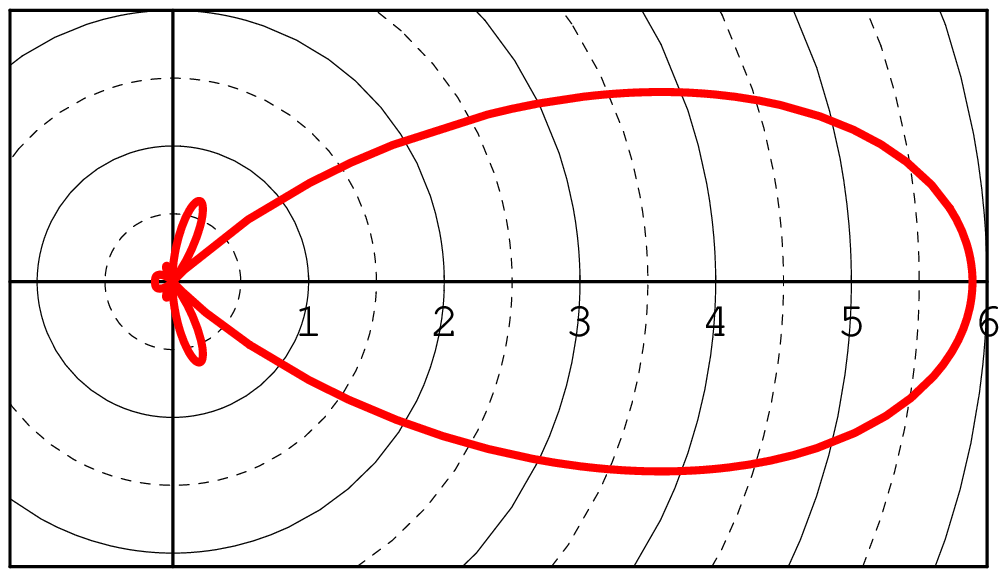}}}}
\put(40,0){\makebox(280,200){
\scalebox{1.0}[1.0]{\includegraphics{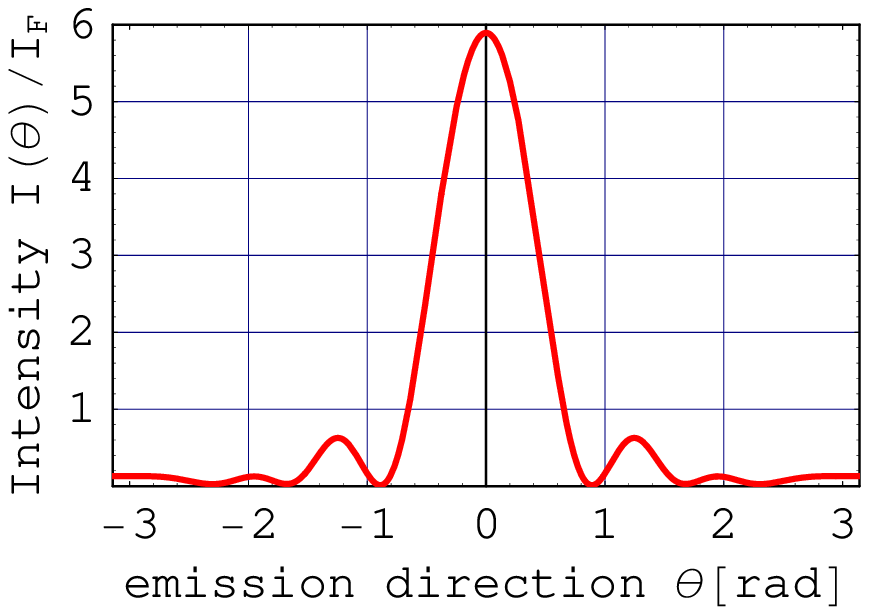}}}}
\end{picture}
\end{center}
\caption{\label{ideal}
Emission pattern of an antenna with negligible losses.
(a) shows the polar plot and (b) shows the rectangular plot.
The antenna parameters are $x_r=-1.4$, $x_a=2$, $\delta=1$
and $\gamma_n=0$.
}
\end{figure}

Fig. \ref{ideal} shows the emission pattern of an antenna with
the above parameters in the limit of negligible absorption
losses ($\gamma_n=0$ for all elements). As expected for a highly
directional antenna, most of the emission is concentrated in a
single lobe around $\theta=0$. In addition, two much smaller
sidelobes exist around angles of $\theta=\pm 1.25$ ($\pm 72^\circ$).
The width of the main lobe can be estimated by using the half
angle $\theta_{1/2}$, defined as the angle between the directions
along which the emission intensity is half of the peak intensity.
In the present case, $I(\theta)/I(0)=1/2$ at $\theta=\pm 0.46$,
so the half angle is $\theta_{1/2}=0.92$ ($53^\circ$).
Compared to the emission of a dipole of amplitude $A_F$, the
forward emission is enhanced by a factor of $I(0)/I_F=5.9$ and
the backward emission is suppressed by $I(\pi)/I_F=0.13$.

\begin{figure}
\begin{center}
\scalebox{0.80}[0.80]{\includegraphics{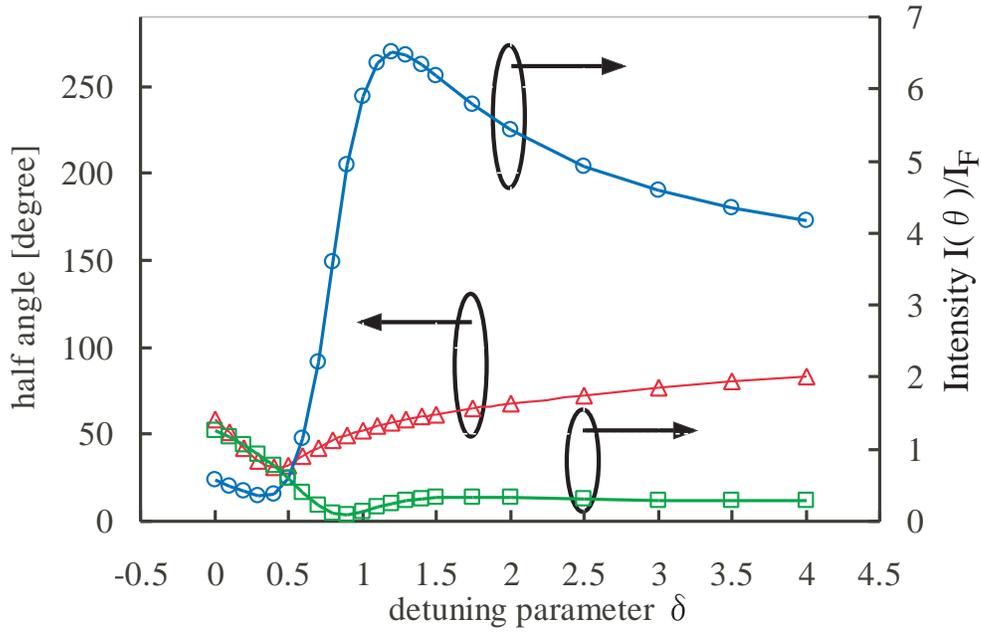}}
\end{center}
\caption{\label{detune} Dependence of antenna emission on
the detuning parameter $\delta$ of the directors. The blue
line shows the enhancement of forward emission, $I(0)/I_F$,
the green line shows the suppression of backward emission,
$I(\pi)/I_F$, and the red line shows the half angle
$\theta_{1/2}$ of the forward emission.
}
\end{figure}

To illustrate the effects of the antenna parameters on the
directionality of the emission, we have determined the
emission patterns for different detuning and director spacings.
Fig. \ref{detune} shows the effect of changing the detuning
parameter $\delta$ on the emission of a five element antenna
with $x_r=-1.4$, $x_a=2$, and $\gamma_n=0$. The importance of
capacitive detuning can be seen clearly in the sharp increase in
forward emission enhancement $I(0)/I_F$, and in the
simultaneous reduction in half angle $\theta_{1/2}$ at a detuning
of about $\delta=0.5$. The forward enhancement reaches a maximum of
about $6.5$ around $\delta=1.2$, and then drops off gradually.
On the whole, the results indicate that it is better to detune too
much than to detune too little, with nearly equally good antenna
patterns for detunings in the range of $\delta=1$ to
$\delta=2.5$.

\begin{figure}
\begin{center}
\scalebox{0.80}[0.80]{\includegraphics{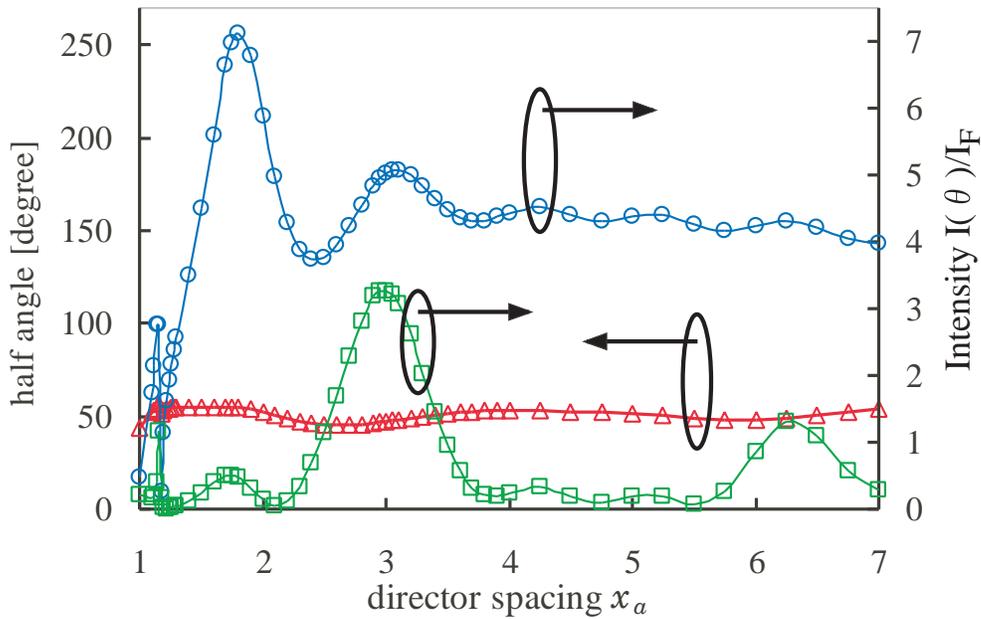}}
\end{center}
\caption{\label{spacing} Dependence of antenna emission on
director spacing $x_a$. The blue
line shows the enhancement of forward emission, $I(0)/I_F$,
the green line shows the suppression of backward emission,
$I(\pi)/I_F$, and the red line shows the half angle
$\theta_{1/2}$ of the forward emission.
}
\end{figure}

Fig. \ref{spacing} shows the effect of changing the antenna spacing
$x_a$ on the emission of a five element antenna
with $x_r=-1.4$, $\delta=1$, and $\gamma_n=0$. As one might expect
for the emission characteristics of an array, the dependence on
array spacing shows some periodic features. However, both forward
and backward emissions oscillate together, so that a spacing of
$x_a=3$ corresponding to half a wavelength causes an enhancement
of backward emission $I(\pi)/I_F$ that is almost as big as
the enhancement of forward emission. These conditions are much better
at $x_a=1.8$, where the forward emission is maximally enhanced by
a factor of about 7.2, whereas the backward emission is suppressed
by about 0.5. A good combination of enhanced forward emission and
strongly suppressed backward emission can be obtained at $x_a=2$
and at $x_a=1.6$, corresponding to the array spacings of
$\lambda/\pi$ and $\lambda/4$ commonly found in radio frequency
antennas.

Interestingly, it is also possible to obtain acceptable antenna
patterns at much wider spacings. In particular, the whole
range of director spacings from $x_a=4$ to $x_a=5.5$ exhibits
very similar radiation patterns with forward enhancements of
about 4 and half angles of about $\theta_{1/2}=0.9$ ($51^\circ$).
It may thus be possible to build antennas with director spacings
of about three quarters of a wavelength that will work even if
the resonant wavelength is shifted considerably.
However, such long antenna structures usually have more sidelobes
at angles between $\theta=0$ and $\theta=\pi$, so
we will focus on the more conventional case of the
$x_a=2$ array in the following.

\section{Emission characteristics of nano-optical
Yagi-Uda antennas with non-negligible absorption losses}
\label{sec:emission}

As we have seen in section \ref{sec:gold}, it may be difficult to
achieve negligible losses in plasmonic nanoparticles.
Specifically, an antenna made of 100 nm long gold nanorods
embedded in glass is expected to have a loss parameter of about
$\gamma=0.5$. In the following, we therefore present the emission
patterns of antennas with non-negligible absorption losses.

\begin{figure}
\begin{center}
\begin{picture}(360,400)
\put(10,380){\makebox(20,20){(a)}}
\put(10,180){\makebox(20,20){(b)}}
\put(40,200){\makebox(280,200){
\scalebox{1.0}[1.0]{\includegraphics{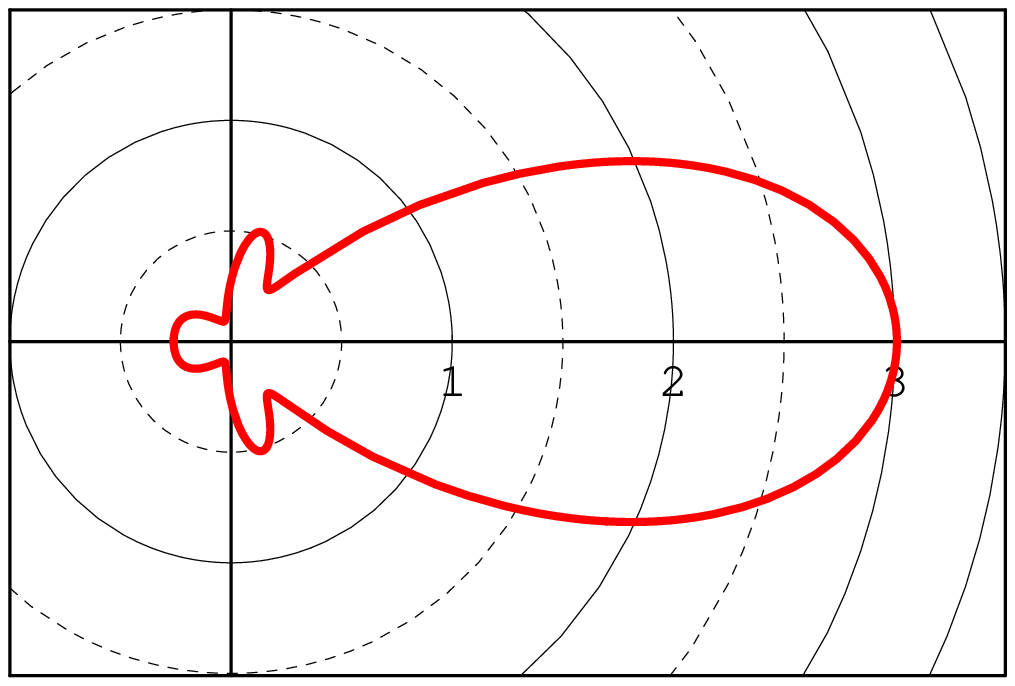}}}}
\put(40,0){\makebox(280,200){
\scalebox{1.0}[1.0]{\includegraphics{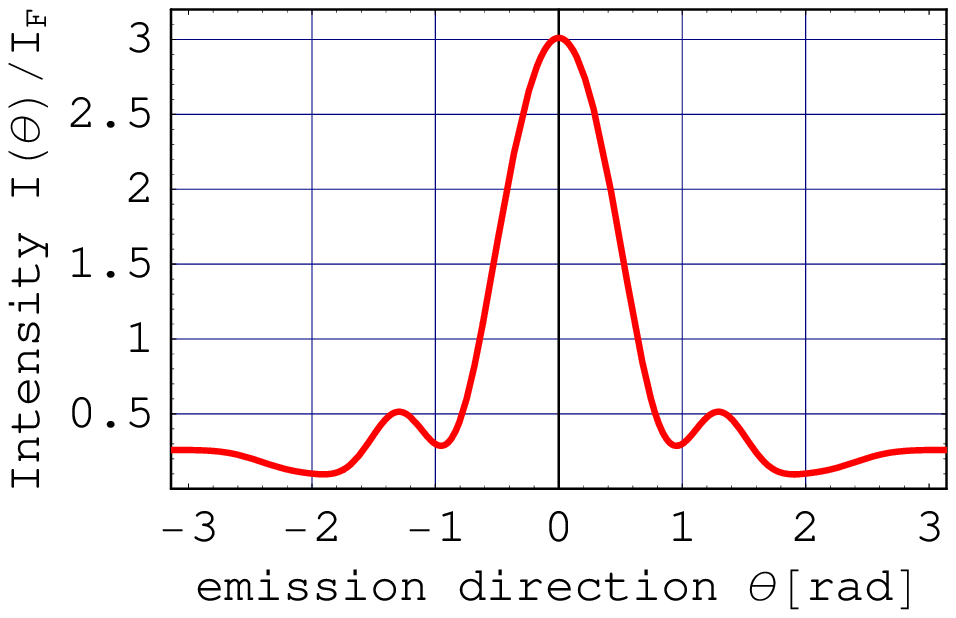}}}}
\end{picture}
\end{center}
\caption{\label{ghalf}Emission pattern of an antenna with
losses equal to $\gamma_n=0.5$ for each element.
(a) shows the polar plot and (b) shows the rectangular plot.
The other antenna parameters are $x_r=-1.4$, $x_a=2$, $\delta=1.5$.}
\end{figure}

\begin{figure}
\begin{center}
\begin{picture}(360,400)
\put(10,380){\makebox(20,20){(a)}}
\put(10,180){\makebox(20,20){(b)}}
\put(40,200){\makebox(280,200){
\scalebox{1.1}[1.1]{\includegraphics{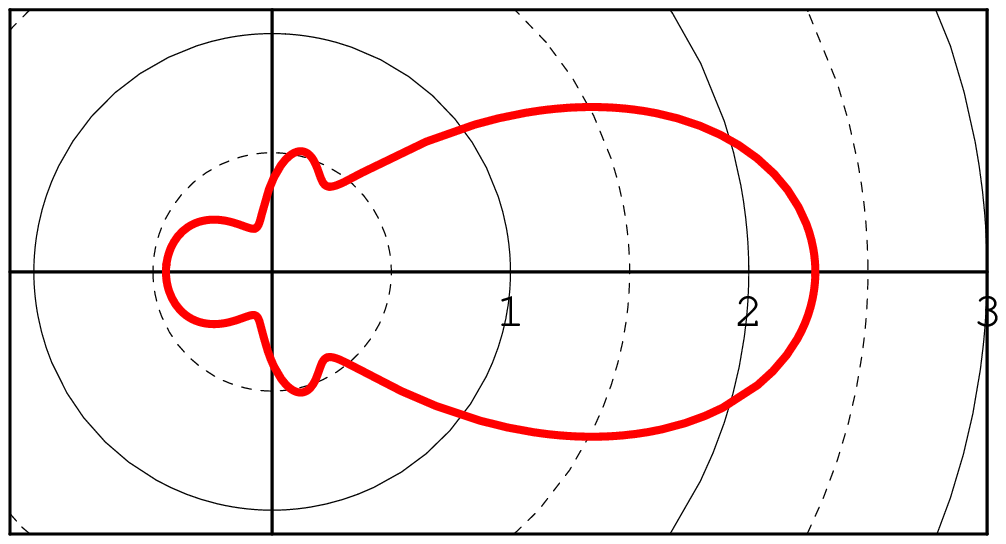}}}}
\put(40,0){\makebox(280,200){
\scalebox{1.0}[1.0]{\includegraphics{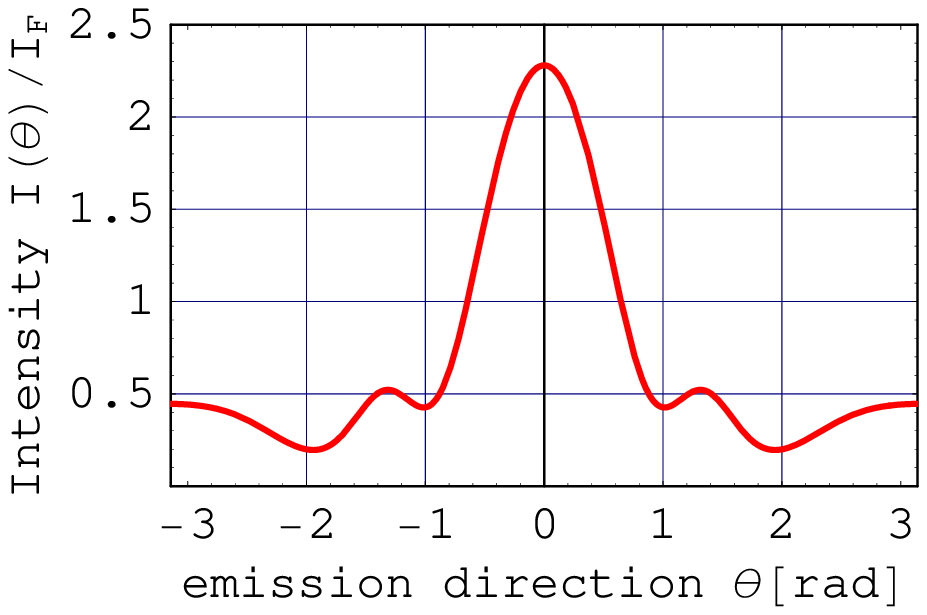}}}}
\end{picture}
\end{center}
\caption{\label{gone}Emission pattern of an antenna with
losses equal to $\gamma_n=1$ for each element.
(a) shows the polar plot and (b) shows the rectangular plot.
The other antenna parameters are $x_r=-1.4$, $x_a=2$, $\delta=2$.}
\end{figure}

Fig. \ref{ghalf} shows the emission pattern for a Yagi-Uda antenna
with $\gamma_n=0.5$ for all elements. The director detuning has been
increased to $\delta=1.5$ to preserve the ratio of
$\delta$ and $1+\gamma$ in eq.(\ref{eq:array}). The array
parameters are $x_r=-1.4$ and $x_a=2$, the same as in
fig. \ref{ideal}. Compared to the loss free case shown in
that figure, the enhancement of forward emission has dropped
by one half to $I(0)/I_F=3$, and the suppression of backward emission
is only $I(\pi)/I_F=0.26$. Still, most of the emission is
concentrated in the main lobe around $\theta=0$, with much
smaller sidelobes around $\theta=\pm 1.29$ ($74^\circ$). However,
the losses have widened the main lobe to a half angle
of $\theta_{1/2}=1.06$ ($61^\circ$), an increase of about 15 \%.
Thus the enhancement of emission in the forward direction is
considerably decreased, but the overall emission pattern is
nearly unchanged. It should therefore be possible to build a
working nano-optical Yagi-Uda antenna from the 100 nm long
gold nanorods discussed at the end of section \ref{sec:gold}.

As losses increase, the emission pattern gradually blurs.
Fig. \ref{gone} shows the emission pattern for a Yagi-Uda antenna
with $\gamma_n=1$ for all elements. The director detuning is
$\delta=2$ and the array spacings are $x_r=-1.4$ and $x_a=2$.
The enhancement of forward emission has now dropped to
$I(0)/I_F=2.3$ and the suppression of backward emission is only
$I(\pi)/I_F=0.45$. The distinction between the main lobe and the
side lobes at $\theta=\pm 1.32$ ($76^\circ$) begins to wash out
and the main lobe has widened to a half angle of $\theta_{1/2}=1.20$ ($69^\circ$), an increase of 30 \% over the loss free case shown
in fig. \ref{ideal}. Interestingly, the emission pattern still has
the characteristic features of the Yagi-Uda antenna, despite the
rather high emission into side lobes.

\section{Conclusions}
\label{sec:concl}

We have shown that a nano-optical Yagi-Uda antenna can be build
from nanoparticles if the absorption losses are sufficiently small
compared to the radiation losses. Specifically, a good emission
pattern can already be obtained at a loss ratio of $\gamma=0.5$
(absorption losses equal to half the radiation losses).
In general, the loss parameter $\gamma$ can be reduced by
making the nanoparticles larger, so the
material properties define a minimal size for each type of
nanoparticle. We have derived the appropriate sizes for gold nanoparticles embedded in a glass matrix, indicating that a
loss parameter of $\gamma=0.5$ can be obtained for elliptical
gold nanorods of about 115 nm in length.

It might be worth noting that a somewhat smaller size limit
can be obtained for silver nanoparticles. For comparison, the
optical response of silver in glass at 705 nm would be described
by $\epsilon_r-1=11.2 + i 0.14$, corresponding to
$\gamma R = 1.12 \times 10^{-3}$, or about one sixth of
the losses in a gold particle. However, the general size dependence
of losses remains unchanged, so the nanoparticles in an antenna
array cannot be made arbitrarily small, regardless of material.

Other design parameters seem to be less critical. In particular,
the working principle of the Yagi-Uda antenna is not sensitive to
small changes in array spacings and in detuning. We therefore
expect nano-optical Yagi-Uda antennas to work even in limits
were the simple dipole approximation used above ceases to apply.

In summary, the absorption losses in nanoparticles need to be
taken into account when constructing a nano-optical Yagi-Uda
antenna. However, the directionality of light emission from
the antenna can be observed even in the presence of non-negligible
losses such as the ones expected in gold nanorods. Our results
thus indicate that a working nano-optical Yagi-Uda antenna
could be realized by an array of 100 nm gold nanorods embedded in
a glass substrate.

\ack
Part of this work has been supported by the
Grant-in-Aid program of the Japanese Society for the
Promotion of Science and the Furukawa technology foundation.


\vspace{1cm}

\begin{thebibliography}{xyz00}

\bibitem{Eng05}
Engheta N, Salandrino A, and Alu A 2005 \PRL {\bf 95} 095504

\bibitem{Mai05}
Maier S A and Atwater H A 2005 \JAP {\bf 98} 011101

\bibitem{Ger00}
Gersen H, Garcia-Parajo M F, Novotny L, Veerman J A, Kuipers L
and van Hulst N F 2000 \PRL {\bf 00} 5312

\bibitem{Gre05}
Greffet J J 2005 {\it Science} {\bf 308} 1561

\bibitem{Sch05}
Schuck P J, Fromm D P, Sundaramurthy A, Kino G S and Moerner W E
2005 \PRL {\bf 94} 017402

\bibitem{Muh05}
M\"uhlschlegel P, Eisler H-J, Martin O J F, Hecht B and
Pohl D W 2005 {\it Science} {\bf 308} 1607

\bibitem{Far05}
Farahani J N, Pohl D W, Eisler H J and Hecht B 2005
\PRL {\bf 95} 017402

\bibitem{Cub06}
Cubukcu E, Kort E A, Crozier K B and Capasso F 2006
{\it Appl. Phys. Lett.} {\bf 89} 093120

\bibitem{Kos06}
Kosako T, Yamashita T, Hofmann H F and Kadoya Y 2006
{\it Extended abstracts of the 67 th autumn meeting of
the Japan Society of Applied Physics, August 28th
to September 1st 2006, Kusatsu, Japan} p~943

\bibitem{Li07}
Li J, Salandrino A and Engheta N 2007 {\it e-print} cond-mat/0703086

\bibitem{Joh72}
Johnson P B and Christy R W (1972) \PR B {\bf 6}, 4370

\bibitem{Sun05}
Sun L, Hao Y, Chien C L and Searson P C 2005 {\it IBM J. Res. \&
Dev.} {\bf 49}, 79

\bibitem{Mar93}
Markel V A 1993 {\it J. Mod. Opt.} {\bf 40} 2281

\bibitem{Zou06}
Zou S and Schatz G C 2006 \NT {\bf 17} 2813

\end{thebibliography}
\end{document}